\documentstyle[preprint,aps,prl]{revtex}

\begin{document}

\draft
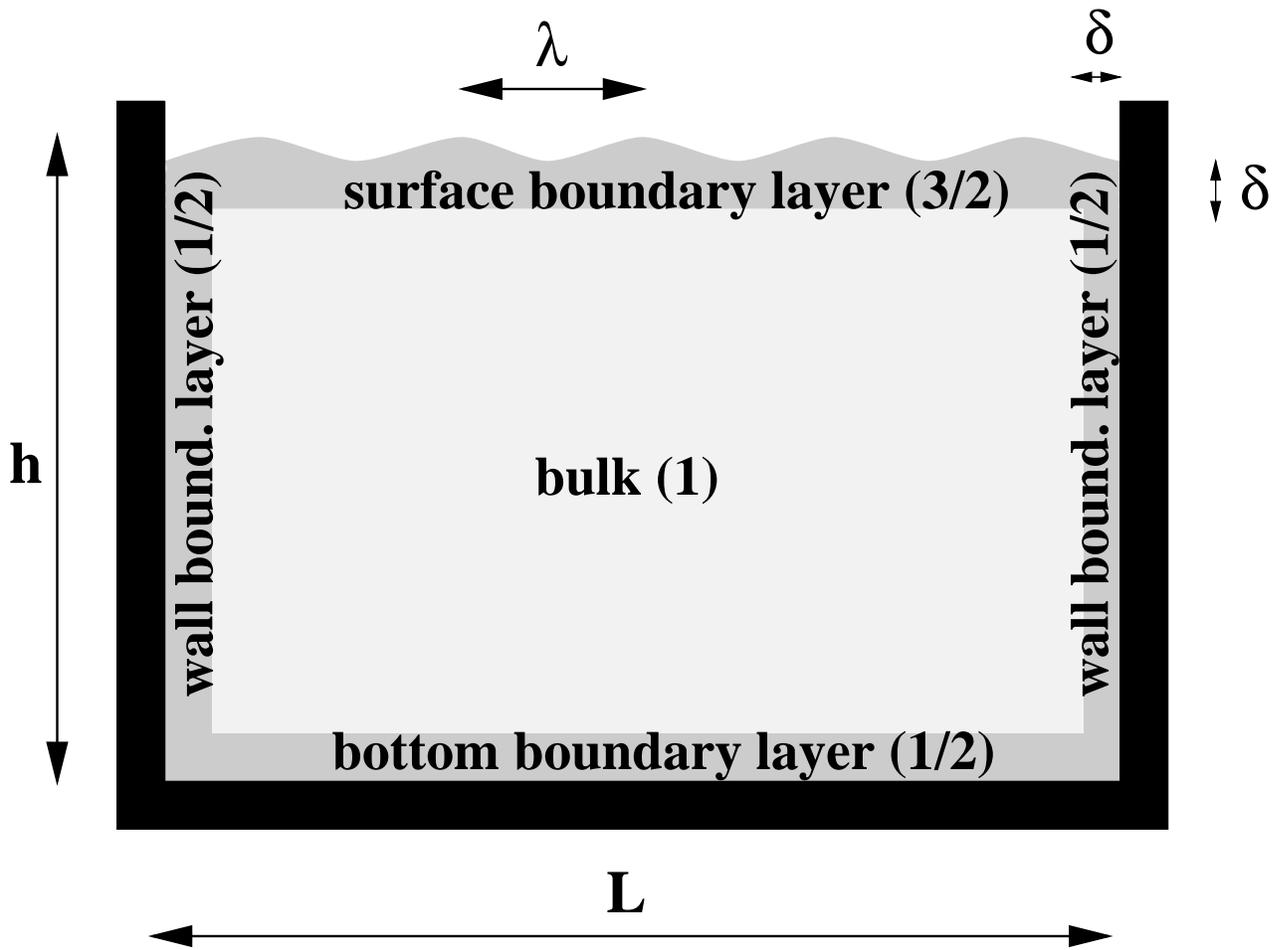
\title{An analytical stability theory for Faraday waves \\ and the observation
of the harmonic surface response}
\author{H.~W.~M\"uller\dag, H.~Wittmer\dag, C.~Wagner\ddag, 
 J.~Albers\ddag, K.~Knorr\ddag}
\address{\dag Institut f\"ur Theoretische Physik, Universit\"at des Saarlandes,
         Postfach 151150, D 66041 Saarbr\"ucken, Germany\\
\ddag Institut 
f\"ur Technische Physik, Universit\"at des Saarlandes,
         Postfach 151150, D 66041 Saarbr\"ucken, Germany}
\maketitle

\begin{abstract}
We present an  analytical stability theory for the onset
of the Faraday instability, applying  over a wide frequency range
between shallow water gravity and deep water capillary waves. For sufficiently
 thin fluid layers 
the surface is predicted to occur in harmonic rather than subharmonic 
resonance with the forcing.  An 
experimental confirmation  of this result is given. 
\end{abstract}
\pacs{PACS: 47.20.Ma, 47.20.Gv, 47.15.Cb}
%

\narrowtext 
The observation of standing waves at the surface of a fluid layer subjected
to a vertical sinusoidal vibration dates back to Faraday \cite{faraday31}. 
For sufficiently 
strong forcing the plane surface undergoes an instabilty giving rise to ordered
patterns of different orientational order \cite{tufillaro89,miles90,gollub91}. 
For purely sinusoidal forcing
spatially periodic patterns of lines, squares 
\cite{douady90,edwards94}, hexagons and  triangles \cite{kumar95} have
been found, even quasiperiodic structures of 8-fold orientational order 
have been observed \cite{christiansen92}.  
Faraday already recognized that the response 
of the surface appears with twice the period of the forcing. 
The first theoretical investigation of the linear stability \cite{benjamin54} 
showed that 
the problem  can be reduced to a 
set of Mathieu oscillators (parametrically driven pendula). However, the
analysis 
takes advantage of the  potential flow approximation which strictly applies 
 to 
ideal (=inviscid) fluids only. Viscous effects are usually treated by
introducing a heuristic damping term in the Mathieu equations \cite{landau87} 
being proportional
to the kinematic viscosity $\nu$. Its strength is  estimated
by evaluating  the viscous energy dissipation in the bulk of the fluid. This
approximation ignores viscous boundary layers which appear 
close to the borders of the fluid. Within these boundary layers the
potential flow assumption breaks down giving rise to vortical  flow, 
associated with additional damping. Kumar and Tuckerman \cite{kumar94} 
were the first 
who performed  a 
linear stability analysis based on the  viscid hydrodynamic field
equations. Their fully numerical analysis uses  Hill's infinite
determinant \cite{nayfeh79} and allows to approximate the critical 
quantities with any 
desired accuracy.  They 
pointed out that the stability problem for viscous fluids cannot rigorously
be reduced to a damped Mathieu oscillator. Beyer and Friedrich \cite{beyer95} 
demonstrated that a systematic treatment
of the viscosity gives rise to a memory term in the Mathieu equation leading
to an integro-differential equation. The integral  gives
rise to additional damping scaling  -- unlike bulk damping -- with
$\nu^{3/2}$. Since Beyer and Friedrich 
used 
idealized free slip boundary conditions  they did 
not catch damping in the bottom boundary layer, which becomes  crucial 
-- as we shall see -- 
for small filling depths. 
Recently Kumar \cite{kumar96} presented an approximative linear analysis 
based on 
a truncation of his  numerical method \cite{kumar94}.
 For small damping he obtained 
 an expression for
the neutral stability curve, which requires a subsequent 
minimization with  respect to the wave number to extract the 
threshold.
Beside of being rather implicit this approach is not systematic
since one does not know to what order of viscosity the obtained result 
is valid \cite{nayfeh79}. Since damping strongly increases as 
the wavelength $\lambda=2 \pi/k$ of the  pattern compares to
the  filling depth $h$,  Kumar \cite{kumar96} points out the possibility of 
observing the
harmonic surface resonance rather than  the subharmonic one. The necessary 
parameter combination is 
beyond of  his analytical approach, that's why he 
provides an example by numerical means.  

In the present paper we give a detailed account of the different
damping mechanisms and develop a perturbative analytical treatment of
the linear stability problem which does not suffer from the earlier 
shortcomings. We provide expressions for the critical
onset acceleration $a_c$ and the critical wave number $k_c$ applying also to
the case of  shallow water waves ($\lambda \simeq h $). 
The 
perturbation analysis is performed for the subharmonic (S) as well as for the 
harmonic (H)
instability, allowing an easy prediction of the bicritical
situation. Finally we present a  Faraday experiment giving the first 
observation of surface waves in {\em harmonic} resonance with the forcing.

We consider a layer of an incompressible fluid of density $\rho$, depth $h$ 
with a free upper surface in contact with air. The layer is subjected to a 
sinusoidal vertical
vibration corresponding to a modulated gravitational acceleration 
$g(t)=g_0+a \cos{2 \Omega t}$ in the comoving frame of reference. 
Control parameters are the strength of the
modulation $a$ and the forcing frequency $f=2 \Omega/(2\pi)$. The free surface 
is 
initially flat at the vertical
coordinate $z=0$. As the  forcing amplitude exceeds a critical thershold
$a_c$ the 
surface
is located at the position $z=\zeta(x,y,t)$, where $x,y$ are the horizontal
coordinates and $t$ is the time. For fluids of viscosity $\nu$ and depth $h$ 
the 
linear dispersion relation \cite{kumar96} for {\em free} (i.e. $a=0$)
surface waves of the form $\zeta \propto \exp{i(kx-\omega t)}$ is
\begin{eqnarray}
\label{dis}
\lefteqn{0= A(k ,\omega) = } \nonumber \\
& &\omega_0^2(k)+ \frac{r(r^4+2r^2+5)\coth{(rkh)}-(1+6r^2+r^4)\tanh{(kh)}}
{r\coth{(rkh)}-\coth{(kh)}}\,\varepsilon^2 - \nonumber \\  
& & \frac{4r(r^2+1)\tanh{(kh)}/(\cosh{(kh)}\cosh{(rkh)})}
{r\coth{(rkh)}-\coth{(kh)}}\,\varepsilon^2\,,
\end{eqnarray}
with $\omega_0^2(k)=\tanh{(kh)}[g_0 k +(\sigma/\rho)k^3]/\Omega^2$, 
$\varepsilon=\nu k^2/|\Omega|$ and $r=\sqrt{1+i\omega/\varepsilon}$. 
The square root with the positive real part is assumed. To shorten 
notation we have nondimensionalized time $t$ and frequencies $\omega,\,
\omega_0$
by the control parameter $\Omega$.
In the ideal fluid limit $\nu \to 0$ Eq.~\ref{dis} reduces to the well
known gravity-capillary dispersion relation 
$\omega^2=\omega_0^2(k)$. The periodic forcing is introduced  by 
replacing  $g_0$ in $\omega_0^2$ by $g(t)$. This parametric driving
couples different  temporal Fourier modes  and
the following linear evolution equation for the Fourier transform
$\hat{\zeta} (\omega)=\int \exp{(-i \omega t)}\, \zeta(t) \, dt$ of the surface
elevation results
\begin{equation}
\label{evol}
A(k,\omega) \hat{\zeta}(\omega) + 
\frac{a\, k \tanh{(kh)}}{2 \Omega^2}\left[\hat{\zeta}(\omega-2)+
\hat{\zeta}(\omega+2)\right]=0\,.
\end{equation}
 It is useful to account for the 
different length scales 
of relevance in the Faraday experiment (see Fig.~\ref{fig1}). 
Intrinsic
are the wavelength $\lambda=2 \pi/k$ of the surface pattern, the 
thickness of the viscous boundary layer $\delta=\sqrt{2 \nu/ |\Omega|}$, 
and the capillary length
$\sqrt{\sigma/ (\rho g_0)}$, where $\sigma$ is the  surface tension. 
Dictated by the geometry are the filling 
depth
$h$ and the lateral dimension of the vessel $L$, which we ignore here by 
assuming $L \to \infty$. 

In order to make analytical progress we confine ourselves 
to the limit of low 
viscosity $\varepsilon \ll 1$, telling that the depth of the
viscous boundary layer is small compared to the wavelength. Furthermore we
assume that $h$ is at least
a few times larger than the thickness of the viscous boundary layer, 
$h/\sqrt{2 \nu/ |\Omega|}\stackrel{\textstyle>}{\sim}3$, giving 
$\coth{(rkh)}\simeq 1$
and $1/\cosh{(r k h)}\simeq 0$. Note that there is no
restriction upon the relation between $h$ and $\lambda$.   
The two simplifications  made are not  very restrictive and include  almost
 all 
recent experiments. Even in the measurements of Edwards and Fauve 
\cite{edwards94} with a
high viscosity water-glycerol mixture 
$\varepsilon$ did not exceed a value of $0.4$. We first expand 
Eq.~\ref{dis} in powers of $1/r=O(\sqrt{\varepsilon})$ giving 
\begin{eqnarray}
\label{approx_1}
A(k,\omega)=\omega_0^2-\omega^2 + i \omega \varepsilon (3+\coth{}^2) +
\frac{\varepsilon^{1/2}\sqrt{\varepsilon+i \omega}^3 }{\sinh{} \cosh{}} 
+\nonumber \\
\varepsilon^{3/2} \sqrt{\varepsilon+i \omega} (-6 \tanh{}+\coth{} + 
\coth{}^3)+ ...,
\end{eqnarray}
where we have abbreviated  $\coth{(k h)}$ by $\coth{}$ {\em etc}. 
This formulation is particularly useful for a  physical interpretation. 
Transformation of Eq.~\ref{approx_1}
into real space yields a damped Mathieu oscillator with integral
contributions \cite{footnote1}
\begin{eqnarray}
\label{extmat}
0=\ddot{\zeta}(t)+\varepsilon (3+coth^2) \dot{\zeta}(t)
+\left[\omega_0^2+\frac{a\,k\,\tanh{}}{\Omega^2}\,
\cos{(2t)}\right]\zeta(t)+ \nonumber \\
\frac{\varepsilon^{1/2}}{\sqrt{\pi}\sinh{}  \cosh{}}\int_{-\infty}^t 
G(t-\tau)(\varepsilon+\partial_\tau)^2
\zeta(\tau) \, d\tau\,+ \nonumber \\
\frac{-6 \tanh{}+\coth{} + \coth{}^3}{\sqrt{\pi}}\varepsilon^{3/2}
\int_{-\infty}^t 
G(t-\tau)(\varepsilon+\partial_\tau)
\zeta(\tau) \, d\tau \,, \nonumber  \\
\end{eqnarray}
where $G(t)=\exp{(-\varepsilon t)}/\sqrt{t}$. Beside the usual (bulk) damping 
($\propto \dot{\zeta})$ the two  integrals also contribute to the 
dissipation. 
This nonlocal temporal behavior  is analogous to the 
propagation  of thermal  waves into a medium which is
heated at its boundary. Here, the moving  surface  
emitts velocity waves into the interior of the fluid giving  rise  
to dissipation which depends on  the  history  of  $\zeta(t)$.
 The first memory integral scales like $O(\nu^{1/2})$ and 
governs the expansion in the shallow water limit  $k h=O(1)$. It is 
associated with damping in the bottom boundary layer and 
 dies out exponentially ($1/(\sinh{kh}\cosh{kh})$) as $kh$  
increases. The  second 
integral (c.f. Ref.~\cite{beyer95}) survives  for  $h  
\to \infty $. It  scales with $\nu^{3/2}$ and is related to the 
dissipation within the surface boundary 
layer.  

To proceed with the linear stability analysis we reexpand the
square roots in Eq.~\ref{approx_1} giving
\begin{equation} 
A(k,\omega)=-\omega^2+X(k,\omega)+\omega_0^2\,, 
\end{equation} 
where all viscous contributions up to $O(\varepsilon^{3/2})$ are collected in  
\begin{eqnarray} 
\label{x}
\lefteqn{X(k,\omega)=\Re(X)+ i\, \Im(X)=} \nonumber\\ 
& & -\varepsilon^{1/2}\frac{ |\omega|^{3/2}}{\sqrt{2}\sinh{}\cosh{}}+
\varepsilon^{3/2}\frac{|\omega|^{1/2}}{2\sqrt{2}}(-15 \tanh{} 
+5 \coth{}+2\coth{}^3)+ \nonumber\\
& & i \, \mbox{sgn}(\omega)\left[\varepsilon^{1/2} \frac{|\omega|^{3/2}}
{\sqrt{2}\sinh{} \cosh{}}+
\varepsilon |\omega|(3+coth^2) +
\varepsilon^{3/2}\frac{|\omega|^{1/2}}{2\sqrt{2}}(-15 \tanh{} +5 
\coth{}+2\coth{}^3)\right]\,.
\end{eqnarray} 
We  first  investigate  the  subharmonic  resonance  of  the   surface  
elevation by introducing
\begin{equation} 
\hat{\zeta}(\omega )=\alpha_1\delta(\omega -1 )+\alpha_2\delta(\omega +1)
+ \hat{\zeta}_1 + ... 
\end{equation} 
in Eq.~\ref{evol}. Assuming   a   small   detuning  
$\omega_0^2-1=O(X)$ the solvability condition for $\hat{\zeta}_1=O(X)$ yields 
the
neutral stability curve 
\begin{equation} 
\left[X(k,1)+(\omega_0^2-1)\right]\left[X(k,-1)+(\omega_0^2-1)\right] 
=\left(\frac{a\, k \tanh{}}{2 \Omega^2}\right)^2\,.
\end{equation} 
The  minimum  of the forcing amplitude  $a$  with  respect  to  $\omega_0^2$  
\cite{footnote2} relates the onset of the subharmonic response to the 
imaginary part of $X$ 
\begin{equation} 
\label{sub}
a_c^{(S)}\simeq\frac{2 \Omega^2}{k_S}\coth{(kh)}\,\Im\left[X(k_S,1)\right], 
\end{equation} 
while the critical wave number $k_S$ results from the dispersion relation 
corrected
by the real part
\begin{equation} 
\label{subk}
\omega_0^2(k_S)\simeq1-\Re\left[X(k_S,1)\right]\,. 
\end{equation} 
For  a  typical  experimental situation  Fig.~\ref{fig2} 
compares our analytical formula Eq.~\ref{sub} with the full numerical 
computation according to the method of 
 Kumar  
and Tuckerman \cite{kumar94}. Excellent agreement is achieved over a wide 
frequency range 
between shallow water gravity and deep water capillary waves. The three
contributions in $\Im(X)$ (Eq.~\ref{x}) are respectively related to damping
in the bottom boundary layer, the bulk, and the surface boundary layer,  
 which become important at low, intermediate, and
higher forcing frequencies. Even the sharp reincrease of the onset amplitude 
at
low frequencies (shallow water limit $kh\simeq1$) is well reproduced.  
Consistence   of   the    perturbation  expansion up to $O(\varepsilon^{3/2})$
   requires    $1/(\sinh{} \cosh{})\leq O(\varepsilon^{1/2})$, which  
reflects a low frequency limit for the validity of Eqs.~\ref{sub},\ref{subk}.
At high excitation frequencies $\Omega$ the validity is limited 
 by the size of the expansion coefficient $\varepsilon$. 

A similar perturbation  expansion  can  
be computed for the first {\em harmonic} stability tongue. We obtain for
the onset 
amplitude 
\begin{equation} 
\label{har}
a_c^{(H)}\simeq\frac{4  \Omega^2}{   k_H}\coth{(k_H h)}
\sqrt{\Im\left[X(k_H,2)\right]}  
\end{equation}  
with the critical wave number $k_H$ determined by 
\begin{equation} 
\omega_0^2(k_H)\simeq 4+\frac{2}{3}\,\Im\left[X(k_H,2)\right]-\Re\left[X(k_H,2)
\right]. 
\end{equation} 
A comparison with  the exact numerical stability analysis (Fig.~\ref{fig2}) 
 shows that the agreement is worse than for the 
subharmonic response. The sharp increase of $a_c^{(S)}$ at low frequencies
leads to an intersection with $a_c^{(H)}$ and thus to a situation where the
harmonic response preempts the subharmonic one. This 
phenomenon, however, is difficult to observe experimentally: The 
corresponding $\Omega$-window is rather narrow as it 
is also limited from below by
the {\em second} subharmonic resonance (see Ref.\cite{kumar96}).
 An additional problem is a technical one and applies to  
 most of the commercial shaker systems being in use for  Faraday experiments:
For the considered 
excitation frequencies ($f\simeq 10 Hz$) 
it is usually the
maximum peak {\em elevation} (but not the maximum {\em force}) which prevents 
the
apparatus from reaching the threshold amplitude $a_c^{(H)}$.
     
In order  to observe the harmonic surface resonance  we  set  up  a 
 Faraday  
experiment.  We  used  a  cylindrical  aluminum  container   of  radius  
$R=45mm$ and depth $h=5mm$. Between $R=35mm$ and the outer edge  the  depth  
continuously decreases to  zero.  This  provides  increasing  damping  
 in order to suppress  meniscus  waves.  
Working fluid is a silicone oil (Dow Corning 200) with  a  viscosity  
of $8.9 mm^2/s$,  surface tension of $19.8 \star  10^{-3}N/m$, and  density of
$0.929g/cm^3$ at a temperature  $T=30^\circ C$. We  use an
electromagnetic shaker V400 (LDS) with a maximum force of $98N$ and an 
indicated maximum 
peak elevation of 8 mm.   
 Waveform generation  as  well  as  
data aquisition  is  performed  by  a  PC. The acceleration is measured by
a piezoelectric device (Bruel and Kjaer 4393). For  
pattern visualization the  container  is lighted  from  the top by  a  
concentric ring ($20cm$ in diameter)  of  50  high  intensity  
light emitting diods (LED) allowing for a stroboscopic illumination. A CCD  
camera located in the middle of the ring observes the pattern from the  
top. The LED's are synchronized  to  the  forcing  and  triggered  by  
either the same or twice the excitation period. Duty cycle as  
well as the relative phase of the illumination  with  respect  to  the  
forcing can be controlled externally. This technique  allows  a  clear  
distiction between the subharmonic  and  harmonic  surface  resonance.  
As explained earlier bottom   damping plays a crucial role for the
appearance of the harmonic Faraday instability; thus    a  depth  to  
wavelength ratio of $kh\stackrel{\textstyle<}{\sim}1$ is required. 
For practical purposes filling  depths much less  
than a millimeter are  unsuitable and difficult to control, we used 
$h\simeq0.8mm$. 
For this parameter combination the stability theory predicts 
the harmonic-subharmonic bicriticality, 
$a_c^{(S)}=a_c^{(H)}$, at $f\simeq 9Hz$,
which  is  
close to the low frequency limit of our apparatus.  
Fig.~\ref{fig3}a  presents a  photograph of a pattern in harmonic 
resonance with
the forcing. Both, the vessel and the surface were oscillating synchronously
with 
$9Hz$. To confirm our observation we show  in  
Fig.~\ref{fig3}b  a subharmonic surface pattern, oscillating again with $9Hz$ 
but
excited with $f=18Hz$. As expected, the wavelength in Fig.~\ref{fig3}a and b 
are the same. Finally
Fig.~\ref{fig3}c depicts a subharmonic pattern driven at $f=10Hz$ exhibiting a 
considerably larger wavelength. 
All patterns are achieved by slowly increasing the 
forcing amplitude $a$ beyond the threshold while keeping  $f$ constant.
The    small  
aspect ratio of our container  (diameter to wavelength ratio) did  not  
allow the observation of nice  ordered  structures.  Thus  redoing  the  
experiment with a larger (and thus havier) container would  indeed  be  
desirable, however, exceeds the specifications of our shaker \cite{footnote3}.

In summary  
we have presented an analytical  theory  for  the  onset  of  
the Faraday instability. The analysis is based on the low  viscosity 
 approximation  and  
assumes a filling depth  larger  than  the thickness of the
 viscous  boundary  layer.  
Almost all  recent  experiments  are  covered  by  these  assumptions.  
Particularly  interesting  is  the  case  of   shallow   water   waves  
($\lambda\simeq h$) for which the harmonic  instability  preempts  the  
subharmonic one.  This  theoretical  prediction  is  confirmed  by  an  
experiment.


{\it Acknowledgements\/} ---  Fruitful discussions with S.~Fauve are
 gratefully
achnowledged. This work is supported by the  Deutsche
Forschungsgemeinschaft.


\begin{figure}
\caption[]{Length scales of relevance: Wavelength of the pattern 
$\lambda=2\pi/k$,
thickness of the viscous boundary layers $\delta=\sqrt{2\nu/\Omega}$, 
depth of the layer $h$, and lateral extension of the container $L$.
Energy dissipation in the various regions scales with different powers  of
the viscosity $\nu$ (exponents are indicated in paranthesis). Waves with a
wavelength larger (smaller) than the capillary length $\rho g_0/\sigma$ 
(not shown in the figure) are
called gravity (capillary) waves.}
\label{fig1}
\end{figure}
\begin{figure}
\caption[]{The critical onset amplitude for the subharmonic (S)
and harmonic (H) Faraday instability as a function of the forcing
frequency $f=\Omega/\pi$. Solid lines correspond to the analytical results 
(Eqs.9 and 11), dashed lines are obtained by an exact numerical
treatment. At low forcing frequencies the harmonic instability 
preempts the subharmonic one. Parameters: $\rho=0.934g/cm^3$, 
$\sigma=20.1\star 10^{-3}N/m$,
 $\nu=10mm^2/s$, $h=1mm$.}
\label{fig2}
\end{figure}
\begin{figure}
\caption[]{Surface patterns as observed in the experiment. (a):
Driving force and surface oscillate synchronously at $f=9Hz$ (harmonic
response).  ; (b) Subharmonic surface response  with $9Hz$ at 
a forcing with
$f=18Hz$; (c) Subharmonic surface oscillation  with $5Hz$ at a forcing with
$f=10Hz$.}
\label{fig3}
\end{figure}


\begin{thebibliography}{99}
%
%
\bibitem{faraday31}M.~Faraday, Philos. Trans. R. Soc. London {\bf 52}, 
319 (1831).
%
\bibitem{tufillaro89}N.~B.~Tufillaro, R.~Ramshankar, and J.~P.~Gollub, 
Phys. Rev. Lett. {\bf 62}, 422 (1989).
%
\bibitem{miles90} J.~W.~Miles and D.~Henderson, Ann. Rev. Fluid Mech. 
{\bf 22},
143 (1990).
%
\bibitem{gollub91}J.~P.Gollub, Physica D {\bf 51}, 501 (1991).
%
%
\bibitem{douady90}S.~Douady, J. Fluid Mech. {\bf 221}, 383 (1990).
%
\bibitem{edwards94}W.~S.~Edwards und S.~Fauve, J. Fluid Mech. {\bf 278}, 123
(1994).
%
\bibitem{kumar95}K.~Kumar, K.~M.~Bajaj, Phys. Rev. E{\bf 52}, R4606 (1995).
%
\bibitem{christiansen92}B.~Christiansen, P.~Alstrom und M.~T.~Levinsen, Phys.
Rev. Lett. {\bf 68}, 2157 (1992).
%
\bibitem{benjamin54}T.~B.~Benjamin und F.~Ursell, Proc. R. Soc. Lond. A 
{\bf 225} 505 (1954).
%

\bibitem{landau87}L.~Landau and E.~M.~Lifshitz, Fluid Mechanics, 
2nd edn. Pergamon Press  (1987).
%
\bibitem{kumar94}K.~Kumar und L.~S.~Tuckerman, J. Fluid Mech. {\bf 279} 49 
(1994).
%
\bibitem{nayfeh79}A.~H.~Nayfeh and D.~T.~Mook, {\it Nonlinear oscillations},
Wiley (1979).
%
%
\bibitem{beyer95}J.~Beyer and R.~Friedrich, Phys. Rev. E{\bf 51}, 1162 (1995). 
%
\bibitem{kumar96}K.~Kumar, Proc. R. Soc. Lond. A{\bf 452}, 1113 (1996). 
%
\bibitem{footnote1} to transform $\sqrt{\varepsilon+i\omega}$  and
$\sqrt{\varepsilon+i\omega}^3$ into time
space use the identity $1/\sqrt{\pi}\int_0^\infty u^{-1/2}\, exp[-(\varepsilon
+ i\omega) u]\,du=
   (\varepsilon+ i\omega)^{-1/2}$.
%
\bibitem{footnote2} Up to $O(X)$ minimization with respect to $k$ and 
$\omega_0^2$ are equivalent.
%
\bibitem{footnote3}To obtain the harmonic surface resonance (Fig.3a)
we already exceeded the  maximum elevation of our shaker by $~50\%$ 
generating a slightly imperfect sinusoidal excitation. However, 
the second harmonic
contribution  to 
the vessel vibration was less than $2\%$ of the amplitude of the $9Hz$-basic
forcing frequency.
%
%
%
%
\end{thebibliography}
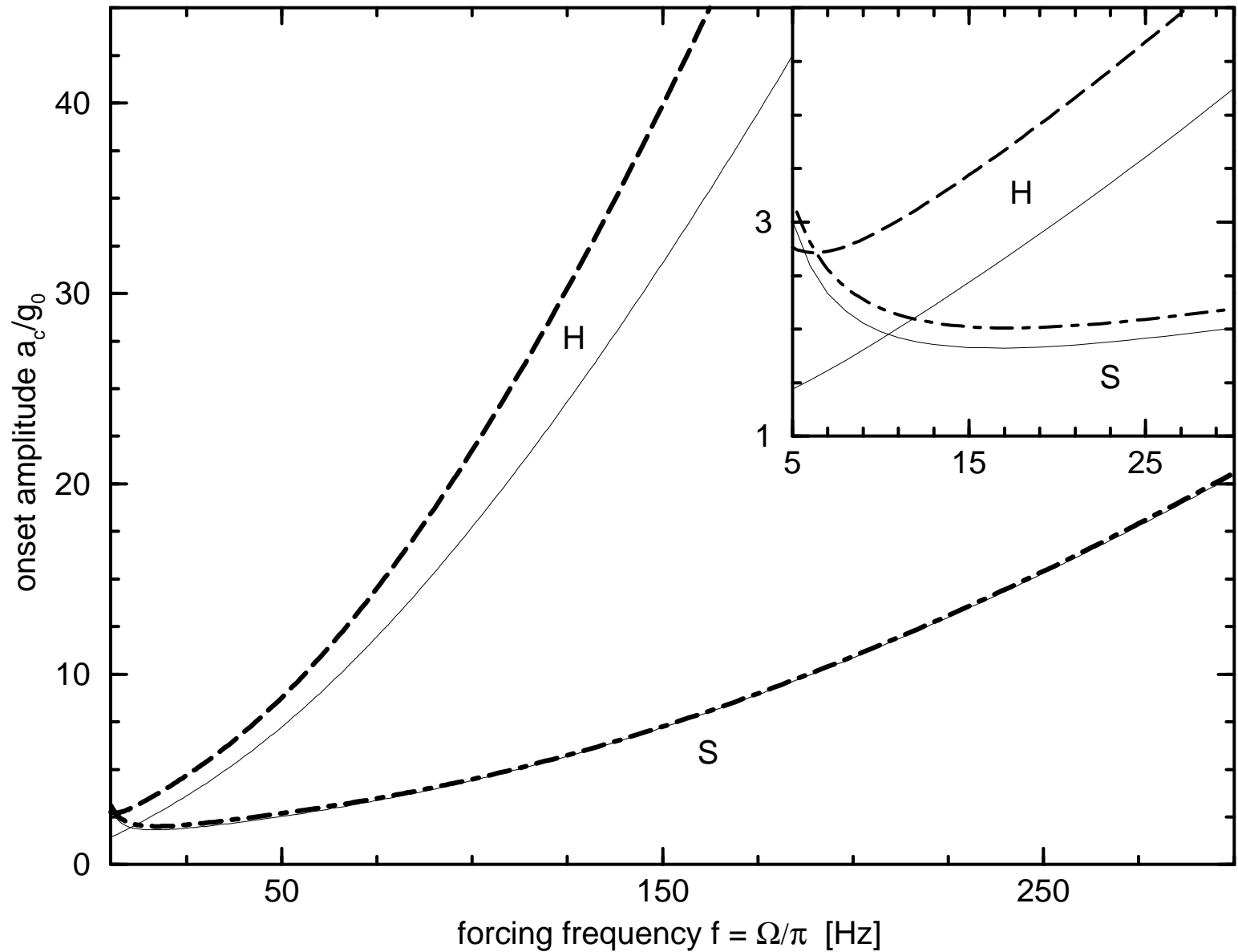
\end{document}